\documentclass[preprint,
eqsecnum,
amsmath,amssymb,
amsmath,amssymb, aps,
]{revtex4}
\usepackage[switch]{lineno}

\usepackage{lineno} 
\usepackage[dvipsnames]{xcolor} 
\usepackage{epsfig}
\usepackage{amsfonts}
\usepackage{amsmath}
\usepackage{wasysym}
\usepackage{amssymb}
\usepackage{bm} 
\usepackage[framemethod=PStricks]{mdframed}
\usepackage{lipsum}
\usepackage{float}

\usepackage{pdflscape}
\usepackage{graphicx}
\usepackage{ulem}

\newcommand{\be}{\begin{equation}}
\newcommand{\ee}{\end{equation}} 
\newcommand{\bea}{\begin{eqnarray}} 
\newcommand{\eea}{\end{eqnarray}}

\usepackage[utf8]{inputenc}
\usepackage{bbm} 				
\usepackage{epstopdf}				
\usepackage{verbatim}    			
\usepackage{sidecap}                
\usepackage{indentfirst}
\usepackage[colorinlistoftodos]{todonotes}

\begin{document}

\title{Top-Down Model of Limescale Formation
in Turbulent Pipe Flows}

\author{L. Moriconi$^1$, T. Nascimento$^2$, B.G.B. de Souza$^2$, and J.B.R. Loureiro$^2$}
\affiliation{$^1$Instituto de F\'\i sica, Universidade Federal do Rio de Janeiro, \\
C.P. 68528, CEP: 21945-970, Rio de Janeiro, RJ, Brazil}
\affiliation{$^2$Programa de Engenharia Mecânica, Coordenação dos Programas de Pós-Graduação em Engenharia, \\
Universidade Federal do Rio de Janeiro, C.P. 68503, CEP: 21945-970, Rio de Janeiro, RJ, Brazil}


\begin{abstract}
We investigate calcium carbonate scale formation at high Reynolds numbers
in a large pipe rig facility. The calcium carbonate solution is produced 
from the injection, at a T-joint inlet, of pH-stabilized sodium carbonate and 
calcium chloride aqueous solutions. A scanning electron 
microscopy analysis of the deposited mass along the pipe indicates that
after an initial transient regime of ion-by-ion crystal growth, calcium 
carbonate scale is dominated by particulate deposition.
While limescale formation in regions that are closer to the pipe's entrance 
can be described as the heterogeneous surface nucleation of calcium and
carbonate ions driven by turbulent diffusion, we rely upon turbophoresis 
phenomenology to devise a peculiarly simple kinetic model of deposition 
at farther downstream regions. Letting $\Phi$ and $R$ be the 
flow rate and the pipe's radius, respectively, 
the mass deposition rates per unit time and unit area are predicted to 
scale as $\Phi^\alpha / R^\beta$ (for certain modeled values of the $\alpha$ 
and $\beta$ parameters) with suggestive support from our experiments.
\end{abstract}

\maketitle


\section{Introduction}

The spontaneous formation of calcium carbonate  (CaCO$_3$) scale, also known as limescale, is an ordinary phenomenon long known to affect the efficiency of water supply networks, boilers and heat exchangers
(Koutsoukos and Kontoyannis 1984; Lin et al. 2020). Limescale formation is also a great source of worry in the oil industry, where it leads, from time to time, to operational issues that are commonly found from the deep production pipelines installed in oil reservoir cores up to topside facilities (Vetter and Farone 1987; Neville 2012; Olajire 2015). Particularly dramatic instances of calcium carbonate scale are those related to the exploration of carbonate reservoirs, which actually hold about 60$\%$ of all of the world oil reserves (Burchette 2012).

Stated in the most synthetic way, calcium carbonate molecules are produced from the binding of calcium and carbonate ions in aqueous solution (Koutsoukos and Kontoyannis 1984), as represented by the following chemical reaction,
\vspace{0.2cm}

\centerline{H$_2$O + CO$_2$ + CaCO$_3$ $\leftrightarrow$ Ca$^{2+}$ + 2 HCO$^{-}_ 3$ .}
\vspace{0.2cm}


\noindent Pressure, temperature or ion concentration variations that come along the lifetime of an 
oil production setup, for instance, eventually drive the above reaction to the left, when the calcium carbonate solution becomes supersaturated and breaks the thermodynamic equilibrium of the once-untouched oil reservoir. Limescale can then grow by heterogenous nucleation of ions at surfaces or by the deposition of bulk pre-existent (nucleated or aggregated) calcium carbonate particles. It is, actually, a matter of great interest to 
understand under what conditions limescale develops through ion-by-ion growth or particulate deposition 
(Broby et al. 2016; Bello 2017).

The immense global budgets applied to the mitigation of calcium carbonate scale formation in all of its manifestations are 
not associated, as a rule, to environment-friendly procedures, to the extent that they are tied to the use of pollutant chemical additives. It is worth mentioning, nevertheless, that remarkable efforts have been done to devise green chemical agents and the implementation of alternative clean anti-scale methods, as the ones based on the use of ultrasound or eletromagnetic devices (MacAdam and Parsons 2004; Kumar 2018; Mazumder 2020). In this connection, it is almost unnecessary to emphasize that progress in the understanding of the physical-chemical basis of limescale formation has a strong potential for innovations of wide social relevance.

Even though the detailed out-of-equilibrium dynamics of calcium carbonate nucleation is still barely understood 
(Gebauer et al. 2008; Wallace et al. 2013; Nielsen et al. 2014; Koishi 2017), its metastable/stable phases can be addressed through thermodynamic modeling or customized experiments (Carino et al. 2017; Cosmo et al. 2019). It is also viable to investigate the formation of limescale (in the presence or absence of chemical additivies) in capillary tubes, with pressures, temperatures, and solution samples which emulate the ones unveiled in realistic situations
(de Souza et al. 2019).

A natural difficulty in weighting the relevance of thermodynamic modeling and the performance of capillary 
flow experiments comes from the fact that in the real world the production of calcium carbonate scale does 
not develop, usually, under quasi-static conditions. Turbulence with a variety of boundary conditions 
and/or sudden variations of thermodynamic parameters provide, actually, the general stage where deposited
limescale is observed. 

In the present study, we avoid thermodynamic and chemical complications, restricting our work to supersaturated 
calcium carbonate aqueous solutions. We focus on the phenomenon of limescale formation in turbulent pipe flows, as realized in experiments performed under ambient temperatures and outlet atmospheric pressure. In order to model the mass deposition rate along the pipe, we take the standpoint, to be concluded from scanning electron microscopy (SEM) images, that the dominant mechanisms of scale formation change along the pipe, ranging from ion crystallization, for some extension after the inlet, to the deposition of calcium carbonate particles at farther downstream regions.

Having in mind the enormous complexity of limescale formation under turbulent conditions, related to the coupling of hydrodynamic and chemical phenomena over a broad range of length scales, we rely on a much simpler top-bottom modeling strategy. A reduced number of hypotheses are put forward, which take into account elementary known facts about near-wall particle-laden flows, combined with feedback experimental inputs.  

This paper is organized as follows. The experimental setup, the experiment preparation and its measurement protocols are detailed in Sec. II. Next, in Sec. III, we collect empirical information associated to the observed limescale formation in the pipe flow experiments and devise, accordingly, elementary models for the prediction of mass deposition rates. We work upon qualitative hints derived from SEM images (Hitachi Tabletop Microscope TM3030), pressure drop measurements and particle size distributions (Malvern Panalytical Mastersizer 3000) taken along the flow. Closer to the inlet region, calcium carbonate deposition is modeled as a turbulent diffusive process of calcium and carbonate ions in the vicinity of the pipe's inner surface. For the far downstream regions, on the other turn, we resort to the phenomenology of turbophoresis phenomena
(Reeks 1983; Young and Leeming 1997; Marchioli and Soldati 2002; Guha 2008; Picano et al. 2009; Sardina et al. 2012) to introduce a system of coupled time evolution equations for the bulk mass densities of two classes of particles, which are characterized by their local Stokes numbers defined in the inner boundary layer region. An appendix provides estimates for the sizes of the calcium carbonate particles that are likely to be captured by the pipe's inner surface, due to attractive van der Waals interactions.

In sec. IV, we report mass deposition rate measurements and discuss them under the light of the present modeling framework. Finally, in Sec. V, we summarize our findings and point out directions of further research.

\section{Experimental Setup and Procedures}

Three reservoirs which are open to the atmosphere (at sea level), with volumes in the range of {\hbox{1.0 - 7.5 m$^3$}}, 
were used to store pH neutral potable water and aqueous solutions of calcium chloride (CaCl$_2$) and sodium bicarbonate (NaHCO$_3$). As a purification protocol, particles with linear sizes smaller than 5 $\mu$m were removed from the water source (provided by the Rio de Janeiro state water company), through the use of carbon activated filters.

As a fundamental procedure to ensure test reproducibility, the aqueous ionic solutions were stirred for about 5 days until their carbon dioxide vapor pressures and pHs reached stable equilibrium values. Only after this preparation step, the turbulent pipe flow was started for the investigation of limescale formation. Table 1 summarizes the physical-chemical properties of the aqueous solutions before they were injected into the pipe.

The ionic solutions were simultaneously pumped through a T-joint inlet (with identical flow rates) into a 70 m long acrylic pipe with inner diameter $D$ of 11 mm. At operating pressures in the range of 1 - 10 bars and environmental temperatures around 300 K, our mixed ionic solutions turned out to be supersaturated (by a far amount) for the precipitation of calcium carbonate particles. The exit flow was directed to a discharge reservoir open to the atmosphere. The essential schematics of the experiment is depicted in Fig. 1.

\begin{table}
\centering
\begin{tabular}{|c|c|c|} \hline
Aqueous Solution of~~ & ~CaCl$_2$~ & ~NaHCO$_3$~ \\ \hline
Concentration (g/l) & 	~7.35 $\pm$ 0.05~ & ~12.6 $\pm$ 0.05~ \\ \hline
~Conductivity (mS/cm)~ & 	~9.73 $\pm$ 0.59~ & ~10.80 $\pm$ 0.38~ \\ \hline
pH & 	~7.48 $\pm$ 0.31~  & ~9.02 $\pm$ 0.26~ \\ \hline
\end{tabular}
\vspace{0.5cm}

\leftline{TABLE 1: Physical-Chemical characterization of the aqueous ionic solutions
used in the turbulent}

\leftline{pipe flow experiments.}

\end{table}
\vspace{0.0cm}

Several pressure taps and points for fluid sampling (for the sake of particle size distribution analyses) were equally spaced along the pipe. In addition, the pipe had twelve {\hbox{14 cm}} lengthwise test sections, distant from each other by approximately 6 m, that could be removed (and reattached) for measurements of the deposited calcium carbonate mass and for the scrutiny of scanning electron microscopy.

\begin{center}
\begin{figure}[ht]
\vspace{0.5cm}

\hspace{0.0cm} \includegraphics[width=0.9\textwidth]{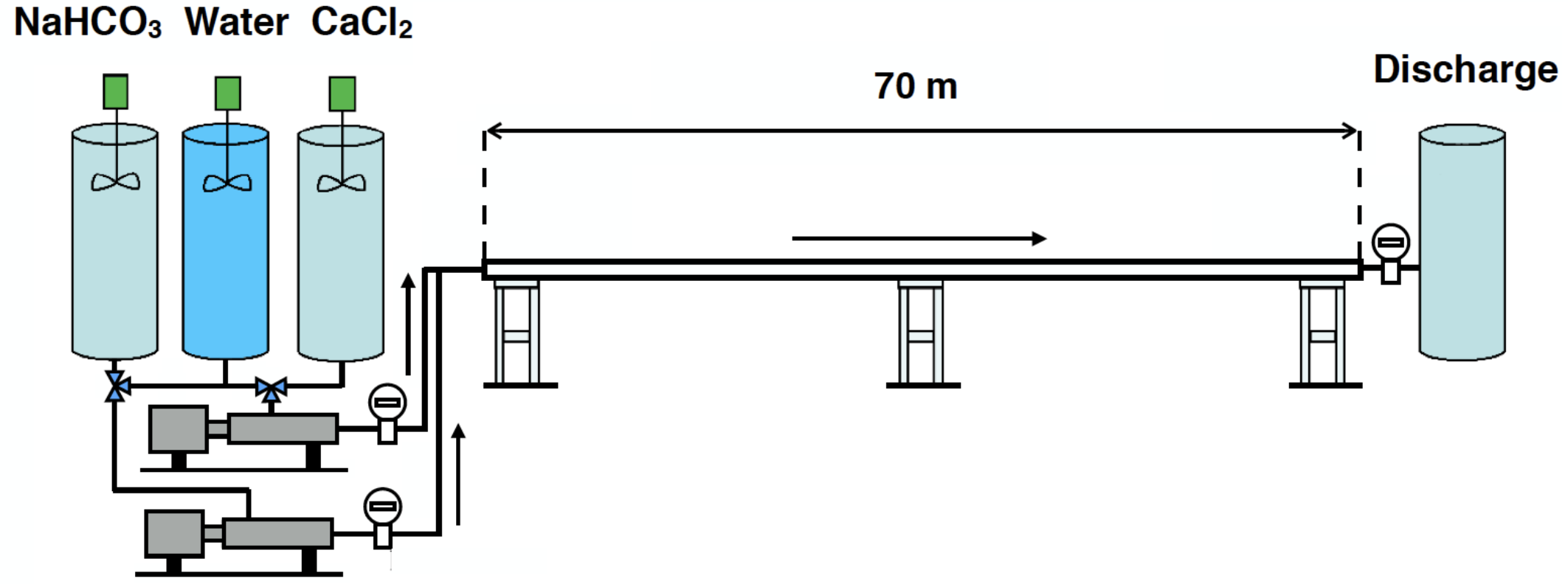}
\vspace{0.5cm}

\caption{Calcium chloride (CaCl$_2$) and sodium bicarbonate (NaHCO$_3$) aqueous solutions are stored in two separate reservoirs. With the help of three-way valves, both of them are initially closed, so that the water reservoir can be used to establish a turbulent pipe flow free of air bubbles. When this is accomplished, the water reservoir is closed and the ionic solutions are pumped into the pipe, without variations of the total flow rate. Not shown in the picture is a fifth reservoir which stores an aqueous solution of acetic acid (10$\%$ v/v), which is employed 
to clean the pipe from the limescale that remains attached to it at the end of the tests.}
\end{figure}
\end{center}

The viscosity of the very dilute solutions of ions and suspended particles was identified, in practice, to the one of water. We have studied limescale formation for five flow rates given by {\hbox{$\Phi = 300, 450, 600, 750,$ and $1000$ l/h}}, corresponding, approximately, to the Reynolds number range $10^4 <$ Re $< 3 \times 10^4$. 


\section{Deposition Model}

We address here a number of phenomenological aspects of limescale formation,
which are particularly related to our experiments and constitute the basic 
ingredients for the modeling considerations carried on in this work. 

It is important to stress, as a remark of methodological nature, that the
data we have collected was not primarily aimed to support any pre-established 
model of ion/particulate deposition. The results of the experimental campaign 
reported below, rather, were intended to provide characterization data, i.e., 
empirical pieces of information ``in search" of a theoretical framework.
\vspace{-0.5cm}

\begin{center}
\begin{figure}[ht]
\hspace{0.0cm} \includegraphics[width=0.9\textwidth]{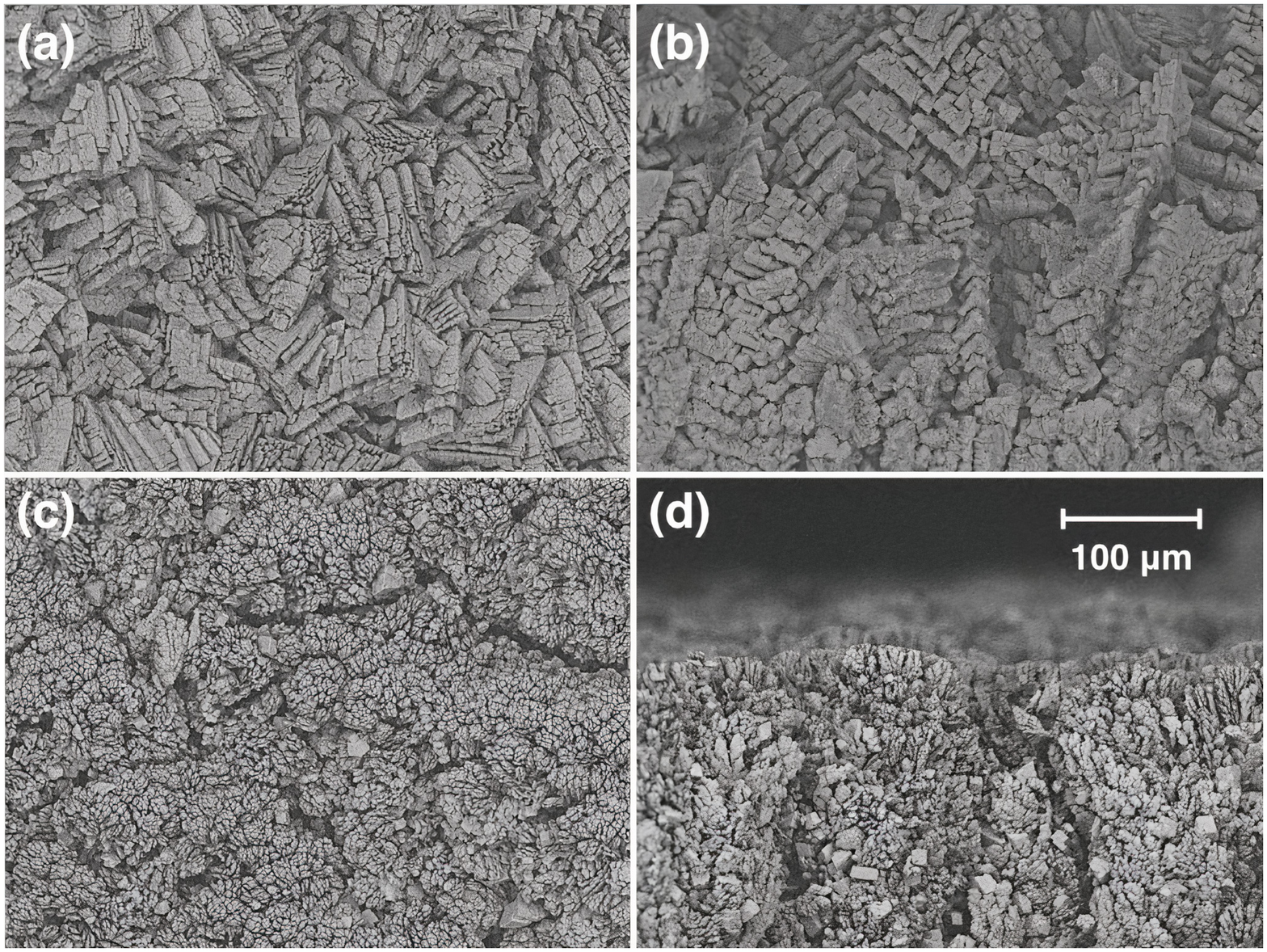}
\hspace{0.0cm} 
\vspace{0.0cm}
\caption{SEM images of calcium carbonate scale in
a turbulent pipe flow ($\Phi$ = 1000 l/h, {\hbox{Re $\simeq 3 \times 10^4$)}} taken 
after 105 mins from the experiment start. 
(a) and (c): SEM top view images;
(b) and (d): SEM side view images. All the pictures have 
the same scale as the one indicated in (d). The snapshot pairs (a,b) and 
(c,d) were taken, respectively at circa 2 and 65 meters away 
from the pipe's inlet. 
Near the inlet, images (a) and (b), the presence of 
faceted calcite crystals indicates limescale formation 
through ion-by-ion crystal growth; far from the inlet,
images (c) and (d), the 
granular structure of limescale suggests that it grows mainly by 
particulate deposition.}
\end{figure}
\end{center}

\subsection{The Nature of the Deposition Process}

Limescale formation is expected to occur, in principle,
through heterogeneous surface nucleation (ion-by-ion interface growth) and/or by the
deposition of calcium carbonate colloidal particles which have sizes 
of a few micrometers. Very small (nanometer-sized) particles, initially 
created through nucleation in the bulk flow, can subsequently grow by ion accretion 
or form larger particles through aggregation.

To resolve the issue of what are the specific scale processes that took place in 
our experiments, we have inspected SEM images of the limescale produced in different 
positions along the pipe. As it is suggested from Figs. 2a and 2b, limescale formation 
is characterized, near the inlet, by the piling and juxtaposition of several broken crystalline pieces of calcium carbonate. The stacked crystal facets visualized in the referred 
images indicate that limescale grows near the pipe's inlet through ion-by-ion deposition. 

Farther downstream, in the last third section of the pipe, the microscopic morphological 
properties of limescale were noticed to be radically changed, as indicated in Figs. 2c and 2d. 
Faceted small crystal pieces of calcium carbonate are much less frequently observed, and 
limescale turns out to be dominated by the aggregation of small-sized ($< 10$ $\mu$m) particles, which form clusters that closely resemble the ones produced in diffusion limited aggregation processes (Witten and Sander 1983).

\subsection{Limescale Roughness}

As it will be clarified in  Secs. III C and III D, a point of particular importance 
in our modeling definitions is the correct assessment of roughness effects on 
mass deposition rates.  Roughness can be usually quantified with the help of 
Moody charts (Çengel and Cimbala 2006), once the friction factor $f$ and the Reynolds number Re are known.
The friction factor is defined, as usual, from the relation (Çengel and Cimbala 2006; Pope 2000)
\be
\frac{d P}{d x} = f \frac{\rho_f}{2}  \frac{\nu^2}{D^3} {\hbox{Re}}^2 \ , \
\label{dpdx}
\ee
for the pressure gradient $dP /dx$, where $x$ is the axial coordinate along the pipe.
Above, $\rho_f$ and $\nu$ are the fluid density and the fluid kinematic viscosity, respectively.

\begin{figure}[ht]
\hspace{0.0cm} \includegraphics[width=0.7\textwidth]{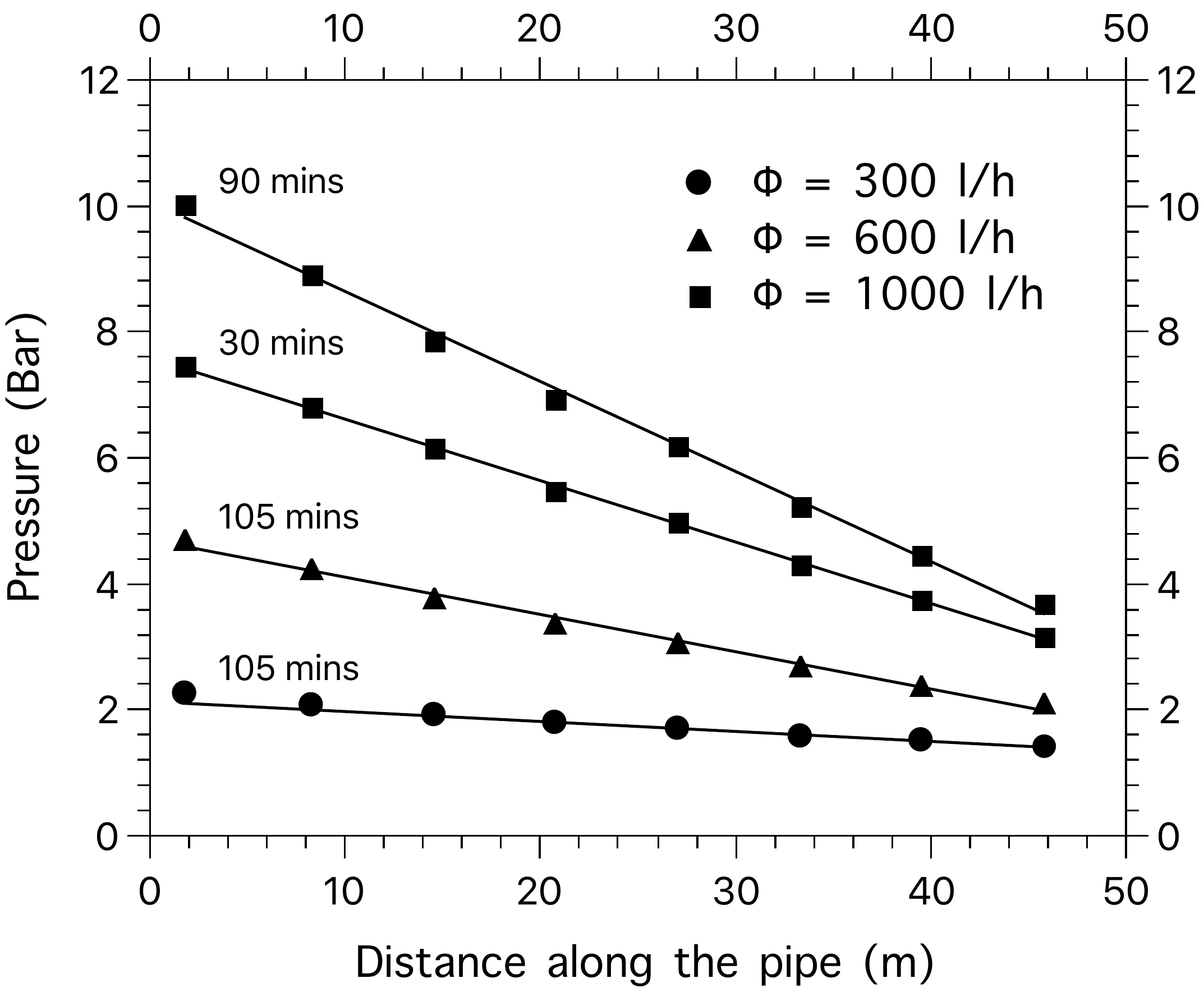}
\vspace{0.2cm}
\caption{A sample of pressure measurements taken along the pipe.
Pressure gradients are approximately constant in space, but time-dependent, due to 
surface roughness growth as deposition proceeds (this is illustrated here for the 
flow rate case {\hbox{$\Phi$ = 1000 l/h}}).}
\end{figure}

Pressure measurements were performed along the pipe in all of the experiments, up to a 
distance of around $45$ meters away from the pipe's inlet. Fig. 3 reports a representative sample of data.
The absolute value of the pressure gradient grows in time as it is pointed out from the data collected 
at different times (30 mins and 90 mins), for the highest Reynolds number case. Roughness growth is, 
in fact, a well-known phenomenon generally associated to the kinematics of evolving interfaces (Barabasi and Stanley 1995).

It turns out, from the evaluation of pressure gradients at different times
and the use of Eq. (\ref{dpdx}), that the friction factor 
evolves, within the time scale of $90$ mins after the flow starts, up to {\hbox{$f \simeq 4.0 \times 10^{-2}$}}, 
regardless the Reynolds number.
We have checked that this estimate is not meaningfully affected if the pressure of 1 bar at 70 meters 
(the pipe's outlet) is added as a common measurement point to all the four datasets in Fig. 3. 

A quick look at a Moody chart tells us, on the other hand, that by taking {\hbox{$f \simeq 4.0 \times 10^{-2}$}} at Re = $10^4$, the friction factor should drop by around $10 \%$ as the Reynolds number increases along the explored range, if the relative pipe roughness were fixed. This means that the pipe relative roughness in our experiments gets larger, for some prescribed evolution time of limescale formation, the larger is the Reynolds number. As a combination of two roughly cancelling effects, thus, the friction factor 
$f$ is approximately Reynolds number independent. It is an interesting issue for additional studies 
to understand to what extent this is a general phenomenon.

\subsection{Limescale Formation by Ion Deposition}

Near the pipe's inlet, calcium and carbonate ions are transported to the wall by turbulent diffusion. 
The rate of mass deposition (mass deposited per unit time and unit area) $\dot \sigma$ can be
written as 
\be
\dot \sigma \propto D_\ast \frac{\delta c}{\ell} \ , \ \label{sigma_dot_ion}
\ee
where $D_\ast \propto \ell v_\ast$ is the turbulent diffusion coefficient and $\ell$ and $v_\ast$  
are the boundary layer scales of length (viscous length) and velocity (friction velocity), 
respectively (Pope 2000). Above, the ion concentration gradient is estimated as $\delta c / \ell$, where $\delta c$
is the ion concentration variation across the viscous sublayer. Taking
$\delta c$ to be flow rate independent (which is a reasonable assumption for the pipe's inlet 
region), we find, from (\ref{sigma_dot_ion}),
\be
\dot \sigma \propto v_\ast = \frac{v_\ast}{U} U \propto \sqrt{f} \frac{\Phi}{R^2} \propto  \frac{\Phi}{R^2} \ , \ \label{sigma_dot_ionb}
\ee
where we have evoked the usual relationship between the dimensionless ratio $v_\ast / U$ and the friction factor $f$ 
(Pope 2000),
\be
f = \frac{1}{8} \left( \frac{v_\ast}{U} \right )^2   \ , \ \label{fpipe}
\ee
and the fact that $f$ is Reynolds number independent in our pipe flow experiments.


\subsection{Limescale Formation and Turbophoresis}

Limescale formation due to particulate deposition, assumed to take place in farther downstream positions along the pipe, requires modeling ideas which are a way more complex than the ones associated to ion-by-ion deposition drove by turbulent diffusion. To start, let us investigate whether significant contributions to deposition may occur through the transport of calcium carbonate particles that behave as tracers.

Since electrostatic or van der Waals interactions have to overcome particle drag forces for deposition to occur, tracer particles will be deposited only if they get close to the walls within atomic length scales (see the discussion carried out in the Appendix). Turbulent diffusion, which should play here a role similar to the one previously addressed for the regime of ion-by-ion deposition, could be conjectured to be the main mechanism for the particulate deposition snapshotted in Figs 2c and 2d. This is, however, an unlikely scenario. As it can be qualitatively noticed from the particle volume distributions provided in Figs. 3a and 3b, there are many more tracer (smaller) particles near the pipe's inlet than elsewhere, while no sign of particulate deposition is found in that region, as already concluded from the inspection of Figs. 2a and 2b. In short, we address henceforth the point that particulate deposition was produced in our experiments through the collisions of larger, inertial particles, with the pipe's inner surface.

\begin{figure}[ht]
\hspace{0.0cm} \includegraphics[width=0.5\textwidth]{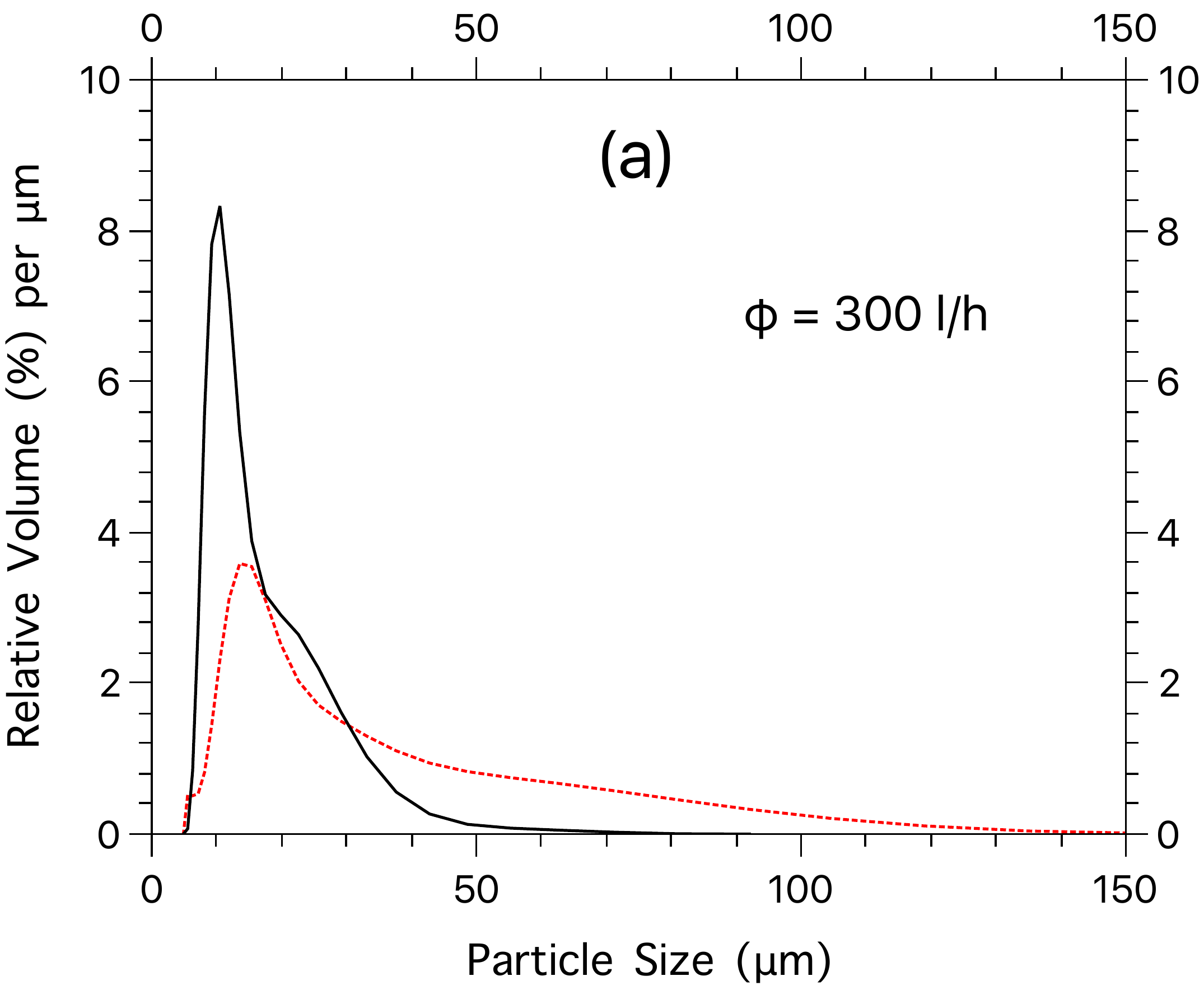}\\
\hspace{0.0cm} \includegraphics[width=0.5\textwidth]{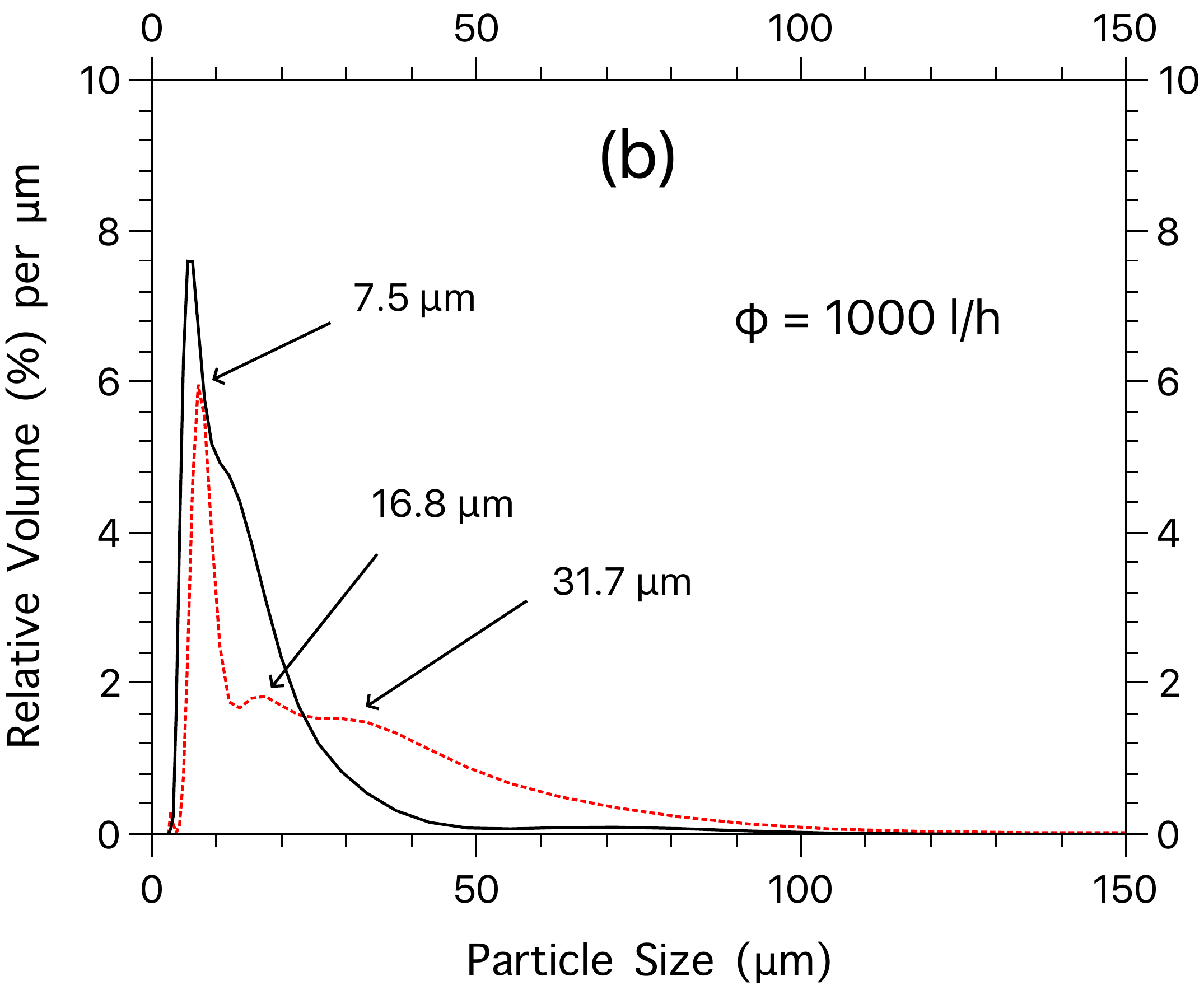}
\caption{Relative volume occupied by suspended particles per particle size variation. Particle size measurements were taken near the pipe's inlet (at distances of {\hbox{2 m}} away from it; black solid lines) and far from the pipe's inlet (at distances of {\hbox{68 m}} away from it; red dotted lines), after the first 30 mins of deposition, for the flow rates of (a) {\hbox{$\Phi$ = 300}} l/h and (b) {\hbox{$\Phi$ = 1000 l/h}}. The particle sizes indicated for the peaks in (b) suggest that pair particle aggregation takes place as the flow develops along the pipe.}
\end{figure}

Once tracers are ruled out as actors of particulate deposition, the phenomenon of turbophoresis
(Reeks 1983; Young and Leeming 1997; Marchioli and Soldati 2002; Guha 2008; Picano et al. 2009; Sardina et al. 2012) comes to mind as soon as we center our attention on the dynamics of inertial particles. In the case of turbulent pipe flows, inertial particles are known to concentrate preferably close to the walls, up to the top of the buffer layer (Picano et al. 2009). It is imperative to devise, thus, estimates of the local Stokes number,
\be
{\hbox{St}}^+ \equiv \frac{1}{18} \frac{\rho_p}{\rho_f} \frac{d^2 v_\ast^2}{\nu^2}  \ , \  \label{st}
\ee
for the transported particles, in order to evaluate their roles in the deposition process. Above, $\rho_p$ and $d$ are, respectively, the particle's mass density and its diameter.

Following the quantitative results established in Picano et al. (2009) and Sardina et al. (2012) on the distribution of 
transported particles in wall turbulence, we adopt the physical picture that particles which 
have Stokes numbers larger than St$^+ \simeq 5$ are the ones which maximize their concentrations in the turbulent boundary layer region $y^+ < 30$. The effective classification of particles as tracers or inertial can be actually put forward from the condition that they have, respectively, Stokes numbers smaller or larger than St$^+ \simeq 5$. Using (\ref{st}), we find that the weakly inertial particles with St$^+ \simeq 5$ have linear sizes around {\hbox{70 $\mu$m and 20 $\mu$m}} for the flow rate cases {\hbox{$\Phi = 300$ l/h}} and {\hbox{$\Phi = 1000$ l/h}}, respectively. 
Referring to Figs. 4a and 4b, we see that those numerical estimates for the sizes of the most concentrated turbophoretic inertial particles are consistent with our view that the deposition of tracers can be neglected in the process of limescale formation. Larger inertial particles are, furthermore, likely to be generated from the pair aggregation of smaller particles, as suggested from the relative volume peaks evidenced in {\hbox{Fig. 4b}}, which are roughly related to each other by powers of two.

It is puzzling, however, that while we expect particles with linear sizes larger or of the order of 20 $\mu$m to be the right candidates for particulate deposition in the flow rate case {\hbox{$\Phi = 1000$ l/h}}, we find, from Figs. 2c and 2d, that the deposited particles have sizes no larger than {\hbox{10 $\mu$m}}. A possible solution of this issue may rely on the hypothesis that particles which collide with the depositing surface are broken up into a number of smaller particles. This is, actually, what happens in the deposition of supercooled large droplets on the surface of airplane wings (Cao and Xin 2020), a largely studied icing phenomenon, relevant for the sake of aircraft flight safety. 

Raising a bridge from the problem of icing on wings to the one of limescale formation in pipe flows is an interesting perspective, whose investigation goes far beyond our present scopes. The main motivation to pursue this goal would be the close analogy that there is between supercooled large droplets of water below the freezing point and metastable calcium carbonate particles in supersaturated solutions.
\vspace{0.2cm}

\leftline{{\it{Reaction-Advection Model of Particulate Deposition}}}
\vspace{0.2cm}


We aim at introducing a model as elementary as possible, in order to account for limescale formation due to particulate deposition in our pipe flow experiments. To begin with, we classify the inertial calcium carbonate particles in the flow into two broad size classes. They are labelled as the classes of ``small" (S) and ``large" (L) particles, operationally defined from the conditions that 
\vspace{0.2cm}

(i) L-particles are the ones which are produced from the aggregation of S-particles 
\vspace{0.2cm}

and that
\vspace{0.2cm}

(ii) S and L-particles have Stokes numbers larger than St$^+ \simeq 5$, the threshold for enhanced particle concentration in the boundary layer range $y^+ < 30$, as suggested from turbophoresis phenomenology.
\vspace{0.2cm}

It is not difficult to show, using (\ref{st}) and mass conservation, that a more quantitative definition of these particle classes can be equivalently given by requiring S and L-particles to have, respectively, Stokes numbers in the ranges $5 \leq$ St$^+ \leq 5 \times 2^{2/3} \simeq 8$ and St$^+ > 8$. These estimates may be subject to further revision, which, however, we do not expect to change our main results, as evidenced from the subsequent arguments.

Let, in self-evident notation, $\rho_s$ and $\rho_l$ be the approximately homogeneous bulk mass densities associated to these two particle classes, at distances $y^+ > 30$ away from the deposition surface. We assume that most of particulate deposition comes from collisions of the more massive L-particles with the pipe walls. The mass deposition rate can then be put forward here as
\be
\dot \sigma = \kappa_l \rho_l v_\ast \ , \  \label{kl}
\ee
where $\kappa_l$ is a dimensionless transport constant and we have taken the friction velocity $v_\ast$ as the natural velocity scale for surface deposition.

\begin{figure}[ht]
\hspace{0.0cm} \includegraphics[width=0.7\textwidth]{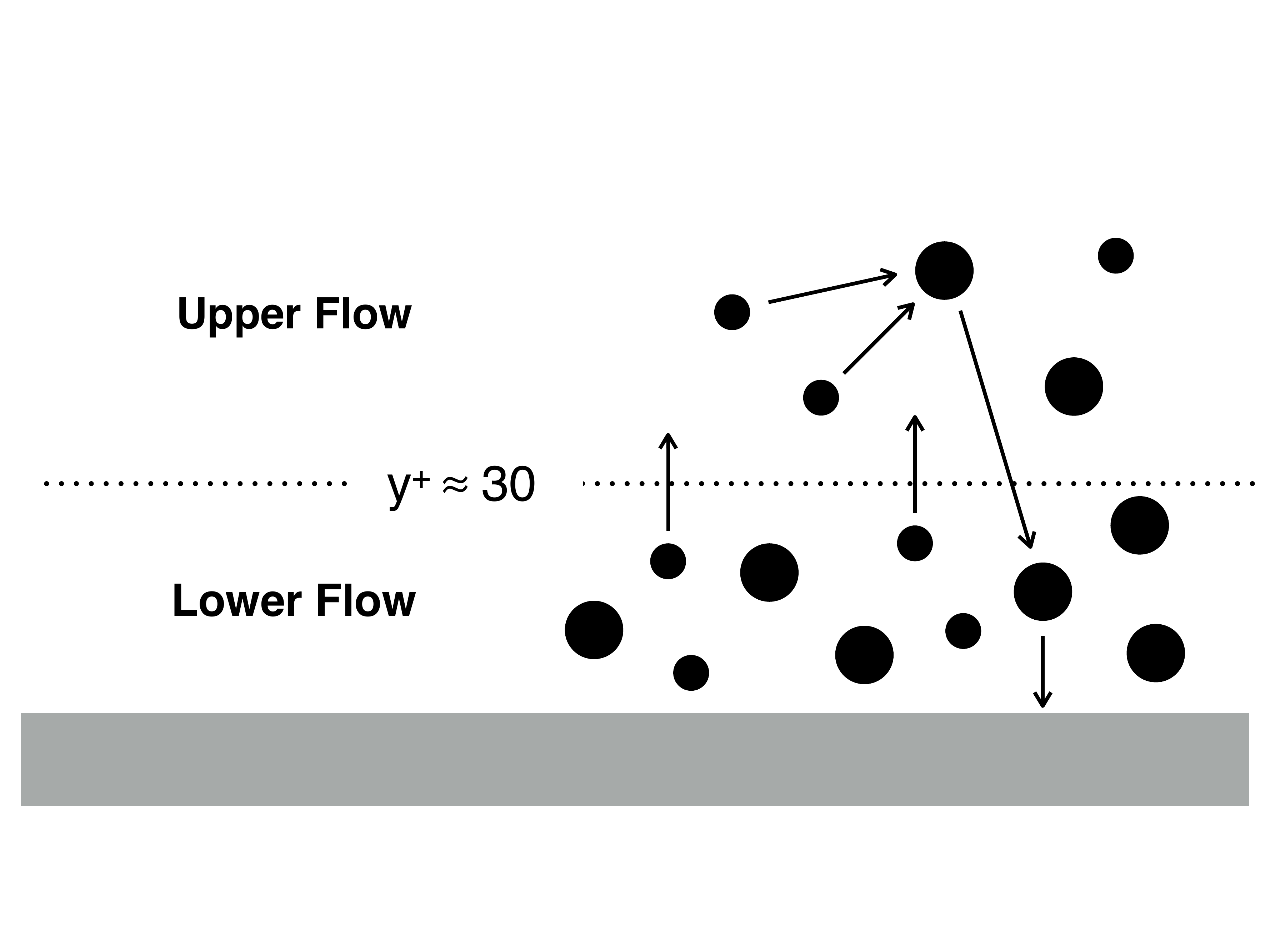}
\caption{Due to turbophoresis (Picano et al. 2009), inertial 
particles tend to accumulate in a lower flow region roughly defined by {\hbox{$y^+ < 30$}}. S and L-particles are schematized in the picture (of course, completely out of scale) as the smaller and larger filled circles, respectively. Particle populations are assumed to be constant in the upper flow, at the expenses of bulk aggregation and unbalanced particle exchanges with the lower flow region. The existence of non-cancelling upward and downward particle fluxes for individual particle classes is ultimately associated to the phenomenon of particulate deposition.}
\end{figure}

If particles did not interact among themselves, we would expect, in the absence of deposition, to have a perfect balance of mass fluxes across the boundary of the turbophoretic region $y^+ < 30$, towards and from upper regions of the flow, for each one of the particle size classes (Picano et al. 2009). However, particles not only aggregate but also break and deposit, so that the detailed balance of incoming and outcoming mass fluxes for both of S and L-particle classes breaks down in the boundary layer. At this point a few important remarks are in order:
\vspace{0.2cm}

{\bf{R1}}. The measured mass deposition rates have been found to approach asymptotic stable values farther downstream from the pipe's inlet, where particulate deposition dominates;
\vspace{0.2cm}

{\bf{R2}}. Particles are intermittently exchanged between the lower ($y^+ < 30$) and the upper ($y^+ > 30$) flow region, due to the drag provided by boundary layer ejections and sweeps (which are then also related to bursts of particle deposition) (Marchioli and Soldati 2002);
\vspace{0.2cm}

{\bf{R3}}. Most of particle aggregation and breakup events are assumed to take place in the upper flow region, 
which contains the vast majority of particles.
\vspace{0.2cm}

Above, {\bf{R1}} imposes a constraint that any conceivable model should comply with; remark {\bf{R2}} implies that both the S and L-particle mass fluxes toward and from the upper flow region can be taken as proportional to $\rho_s v_\ast$ and $\rho_l v_\ast$, respectively; assumption {\bf{R3}}, on its turn, indicates that variations of the bulk mass densities $\rho_s$ and $\rho_l$ associated to aggregation and breakup phenomena can be accounted entirely from balance equations devised for the upper flow domain.

We emphasize, additionally, that as far as L-particles dominate mass deposition, and both $\rho_s$ and $\rho_l$ are asymptotically constant at large distances from the pipe's inlet, in view of {\bf{R1}}, breakup processes can be neglected {\it{vis-à-vis}} the role of S-particle aggregation in the production of bulk L-particles. It is clear, then, that asymptotic plateaus for $\rho_s$ and $\rho_l$ are only possible if the mass of S-particles that is lost in the bulk due to aggregation is compensated through a net flux of S-particles from the inner flow region (which acts like a particle reservoir). This physical picture of deposition is illustrated 
in Fig. 5.

Since our experiments were carried out for dilute particle suspensions, the mass density of L-particles is expected to increase, on the grounds of the binary aggregation of bulk {\hbox{S-particles}}, with rate $\gamma \rho_s^2$, where $\gamma$ is some reaction constant. We can also model the net mass flux of {\hbox{S-particles}} (mass per unit area per unit time) from the lower to the upper flow region as $\kappa_s \rho_s v_\ast$, where $\kappa_s$ is a dimensionless coefficient analogous to the one introduced in Eq. (\ref{kl}). It follows immediately that along a pipe's lengthwise section of extension $\Delta$, the bulk mass of {\hbox{S-particles}} would increase with rate $2 \pi R \Delta \kappa_s \rho_s v_\ast$, if aggregation/breakup effects were neglected. As a consequence, the mass density production rate for the bulk S-particles, neglecting again mass variations from aggregation or breakup processes, is estimated as $2 \pi R \Delta \kappa_s \rho_s v_\ast / \pi R^2 \Delta = 2 \kappa_s \rho_s v_\ast/ R$. In the same fashion, it turns out that the (negative) mass density production rate for the {\hbox{L-particles}}, if restricted to the process of mass transport to the lower flow region (and subsequent deposition) amounts to just $- 2 \kappa_l \rho_l v_\ast / R$.

Putting together the phenomenological points discussed so far, we propose to describe the space-time dependence of the bulk densities 
$\rho_s = \rho_s(x,t)$ and $\rho_l = \rho_l(x,t)$ along the pipe by means of the following minimalist pair of kinetic differential 
equations,
\bea
&&\frac{d \rho_s}{dt} = - \gamma \rho_s^2 + 2 \frac{\kappa_s v_\ast}{R} \rho_s \ , \  \label{dotrhos} \\
&&\frac{d \rho_l}{dt} = \gamma \rho_s^2 - 2 \frac{\kappa_l v_\ast}{R} \rho_l \ , \   \label{dotrhol}
\eea
where $d/dt = \partial_t  + U \partial_x$ is the material derivative for a flow with (effective) advective bulk velocity $U$. Eqs. (\ref{dotrhos}) and (\ref{dotrhol}) can be interpreted as a reduced formulation of population balance dynamics (Ramkrishna 2000) for the dispersed particulate phase.

Far enough from the pipe's inlet, both $d \rho_s / dt$ and $d \rho_l / dt$ vanish, so that
\be
\kappa_s \rho_s = \kappa_l \rho_l \ , \  \label{klks}
\ee
as it can be inferred from (\ref{dotrhos}) and (\ref{dotrhol}).
Once most of the depositing mass is supposed to belong to the L-particle class, we have $\rho_l \gg \rho_s$. Eq. (\ref{klks}) leads, thus, to 
$\kappa_s \gg \kappa_l$. This latter inequality implies that $\rho_s$ relax to its equilibrium value $\bar \rho_s$ 
much faster than $\rho_l$ does. Taking $d \rho_s / dt = 0$ in Eq. (\ref{dotrhos}) and solving for $\rho_s$, it follows that
\be
\bar \rho_s = \frac{ 2 \kappa_s v_\ast}{\gamma R} \ . \ \label{barrhos}
\ee
Replacing $\rho_s$ in the first term of the right-hand side of (\ref{dotrhol}) by (\ref{barrhos}), and working in the stationary 
regime where $\rho_l$ becomes time-independent, Eq. (\ref{dotrhol}) becomes
\be
U \frac{d\rho_l}{dx}  = \frac{4 \kappa_s^2}{\gamma R^2} v_\ast^2 - \frac{2 \kappa_l v_\ast}{R} \rho_l \ . \   \label{drhodt}
\ee
Eq. (\ref{drhodt}) can be straightforwardly solved to give
\be
\rho_l(x) = \rho_0 \exp \left ( - \frac{2 \kappa_l v_\ast}{RU} x \right ) +
\frac{2 \kappa_s^2}{\gamma \kappa_l R} v_\ast \left [
1- \exp \left ( - \frac{2 \kappa_l v_\ast}{RU} x \right ) \right ] \ , \  \label{rholx}
\ee
where $\rho_0 = \rho_l(0)$ is the arbitrary initial condition for the density field $\rho_l(x)$.

It is interesting to express, again, the friction velocity $v_\ast$ in Eq. (\ref{rholx}) as a function of the
friction factor $f$, Eq. (\ref{fpipe}). 
We also introduce in the modeling a friction factor exponent $\alpha$ 
which relates the friction factor to the Reynolds number, as
\be
f \equiv f_0 Re^{-\alpha} = f_0 
\left (\frac{\pi R \nu}{2} \right )^\alpha \Phi^{-\alpha} \ . \ \label{fRe}
\ee
In consonance with the discussion of Sec. III B, we take $\alpha = 0$ in practice. 
Note, in passing, that the standard Blasius relation is given by (\ref{fRe}) with $\alpha = 1/4$
(Pope 2000).

Collecting Eqs. (\ref{fpipe}), (\ref{kl}), (\ref{rholx}), and (\ref{fRe}), the mass deposition rate  
may be written, now, as the position dependent function,
\be
\dot \sigma (x) = a \exp \left ( - b \Phi^{-\alpha/2} x \right ) 
+ c  \Phi^{2-\alpha} \left [
1- \exp \left ( - b \Phi^{-\alpha/2} x \right ) \right ] \ , \  \label{sigma_dot}
\ee
where
\bea 
&&a = \kappa_l \rho_0 v_\ast \ , \ \label{par_a}  \\ 
&&b = 4 \sqrt{2 f_0} \kappa_l \left ( \frac{\pi \nu}{2} \right ) ^{\alpha/2} R^{\alpha/2-1} \ , \ \label{par_b} \\
&&c = \frac{16}{\pi^2} \frac{ \kappa_s^2}{\gamma} f_0 \left ( \frac{\pi \nu}{2} \right ) ^{\alpha} R^{\alpha-5 }\ . \ \label{par_c}
\eea
It could be pointed out that there is a somewhat large room for reasonable phenomenological fittings of mass deposition rates from the use of Eq. (\ref{sigma_dot}), since there are three independent unknown parameters in the model ($a$, $b$, and $c$). However, as argued in the next section, we develop, as a more restrictive test of the model's consistency, a systematic procedure of analysis which considerably reduces the freedom in the choice of fitting parameters.

\section{Results}

Our task now is to discuss the experimental validation of the modeling expressions (\ref{sigma_dot_ion}) 
and (\ref{sigma_dot}) for the mass deposition rates. We start with the latter.

\begin{figure}[b]
\hspace{0.0cm} \includegraphics[width=0.7\textwidth]{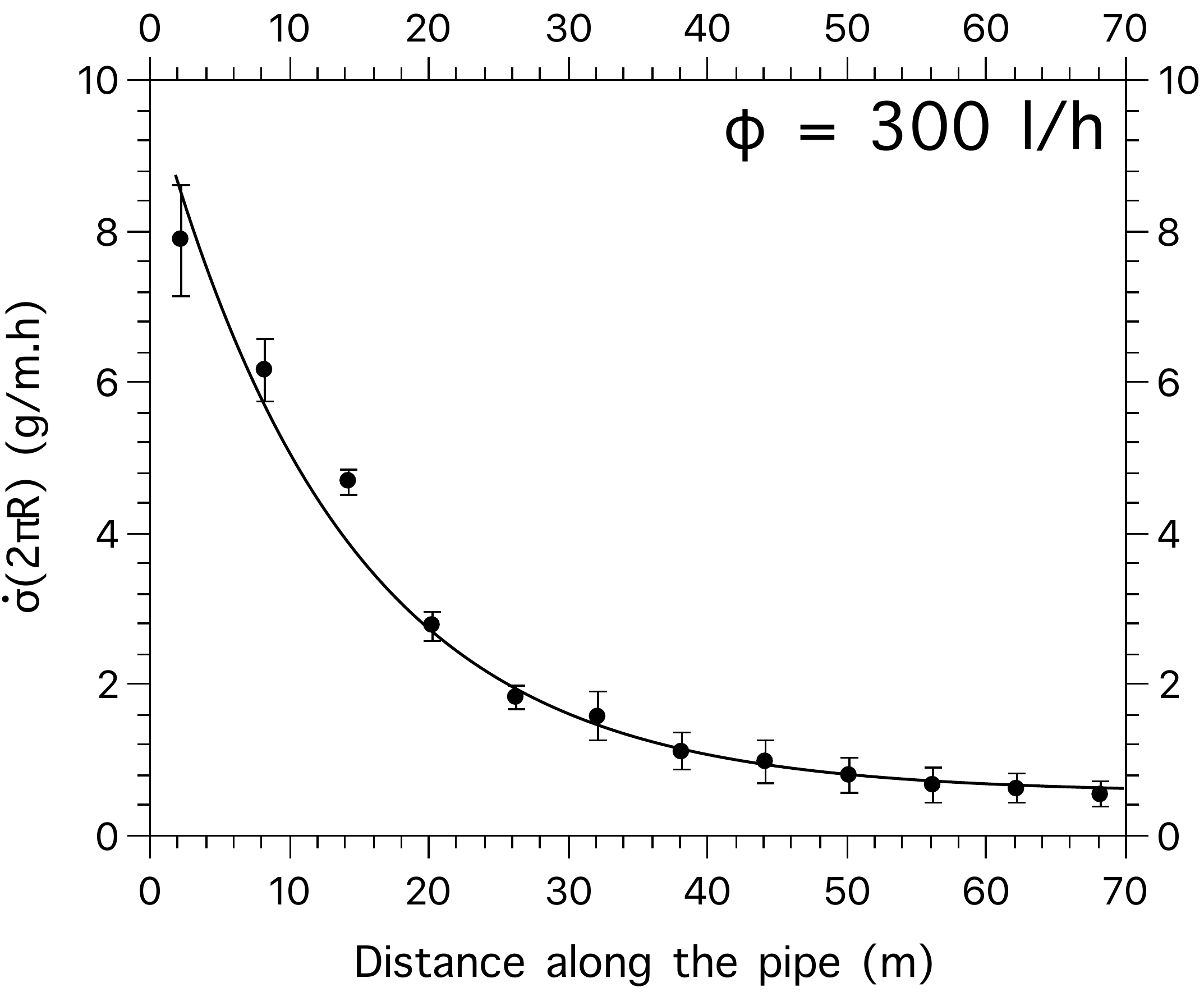}
\caption{The mass deposition rate is here modeled for the flow rate case $\Phi = 300$ l/h,
with the help of Eq. (\ref{sigma_dot}), where we take $\alpha =0$. The error bars are
estimated from three independent experimental runs. The parameter $c$
is empirically obtained from the farthest measurement of the mass deposition rate along the 
pipe, having in mind the asymptotic expression $\dot \sigma = c \Phi^2$. 
Parameters $a$ and $b$ are subsequently found from the best fit of (\ref{sigma_dot}) 
to the measurements taken after the first three data points from left to right. 
We get $a \times 2 \pi R = 10.0$ g/m$\cdot$h, {\hbox{$b=7.3 \times 10^{-2} $ m$^{-1}$ and 
$c \times 2 \pi R = 6.1 \times 10^{-6}$ g/m$\cdot$h.}}} 
\end{figure}

It is interesting to note that the $b$ and $c$ parameters in (\ref{par_b}) and (\ref{par_c}) do not explicity 
depend on the flow rate $\Phi$ or on the initial density parameter $\rho_0$, as in (\ref{par_a}). It is likely, thus, that $b$ and $c$ depend {\it{only}} on the chemical concentrations of calcium and carbonate ions (and of carbon dioxide, as well) injected into the pipe flow, which were approximately the same for all the investigated flow rate cases. Eq. (\ref{sigma_dot}), furthermore, implies that for $\alpha = 0$ the mass deposition rate approaches, far from the pipe's inlet, the asymptotic value 
\be
\dot \sigma_\infty = c \Phi^2 \propto \frac{\Phi^2}{R^5} \ . \ \label{sigma_infty}
\ee
The above simple observations can be used to introduce a consistent strategy for the definition of the open phenomenological parameters in (\ref{sigma_dot}), as it follows.

\begin{figure}[b]
\hspace{0.0cm} \includegraphics[width=0.7\textwidth]{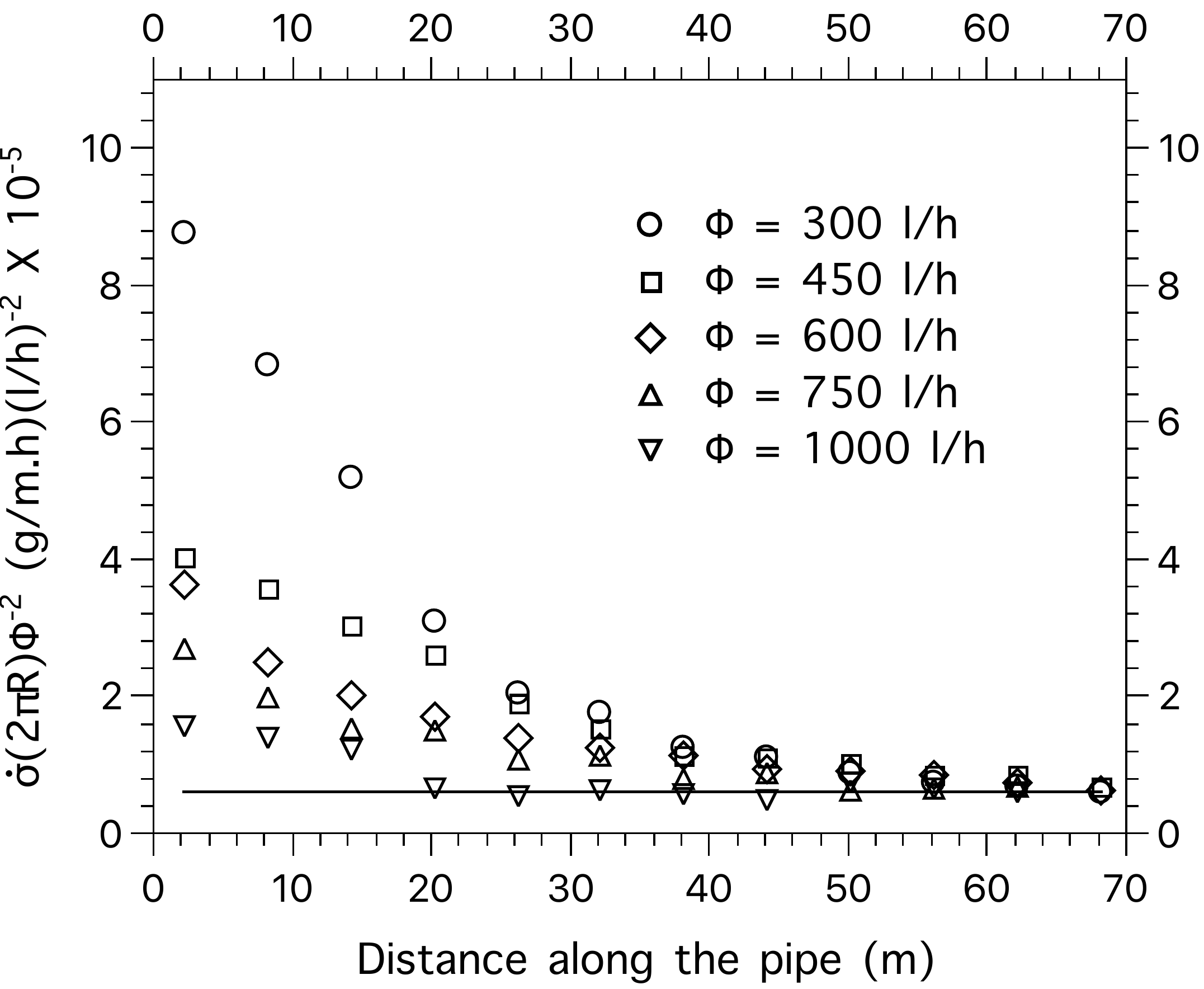}
\vspace{0.0cm}
\caption{Data collapse is achieved for all the measured mass deposition rates at farther downstream positions
along the pipe, if they are normalized by the square of the flow rate. The horizontal solid line follows from the 
asymptotic mass deposition rate, $\dot \sigma_\infty = c \Phi^2$, where the numerical value of $c$ is taken from the 
previous analysis for the flow rate case {\hbox{$\Phi = 300$ l/h}} {\hbox{(see the caption of Fig. 6).}}} 
\label{}
\end{figure}

\vspace{-0.0cm}

\begin{center}
\begin{figure}[b]
\hspace{0.0cm} \includegraphics[width=0.95\textwidth]{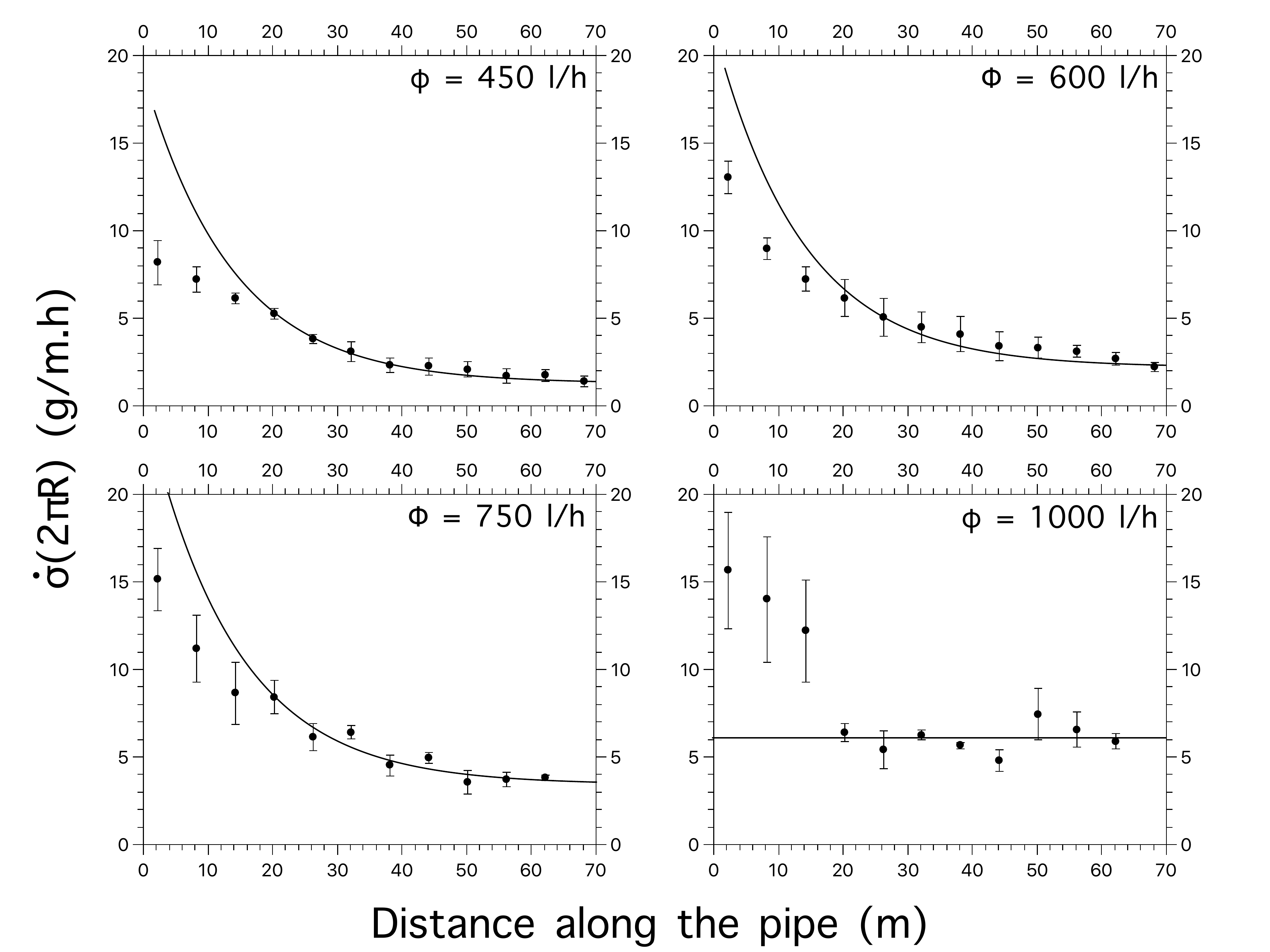}
\vspace{0.0cm}
\caption{This panel shows mass deposition rates for the flow rate cases besides $\Phi = 300$ l/h, 
modeled through Eq. (\ref{sigma_dot}), where $\alpha =0$, $b$ and $c$ have the same values as the ones reported in the caption of Fig. 6, and $a$ is the only free parameter for data fitting (obtained, again, from discarding the first three points from left to right in each dataset). We get, in units of g/m$\cdot$h, $a \times 2 \pi R = 19.3$ for $\Phi = 450$ l/h; $a \times 2 \pi R =21.8$ for $\Phi = 600$ l/h; $a \times 2 \pi R = 25.5$ for $\Phi = 750$ l/h. Due to the much larger error bars for the flow rate case $\Phi = 1000$ l/h, we did not attempt to determine $a$; therefore, we just plot, for this instance, the horizontal line associated to the asymptotic mass deposition rate $\dot \sigma = c \Phi^2$. Error bars were estimated from the number of independent experimental runs for each flow rate case: three runs for $\Phi = 450$ l/h;  five runs for $\Phi = 600$ l/h; six runs for $\Phi = 750$ l/h;  two runs for $\Phi = 1000$ l/h.} 
\label{}
\end{figure}
\end{center}
\vspace{-1.4cm}

\begin{figure}[b]
\hspace{0.0cm} \includegraphics[width=0.7\textwidth]{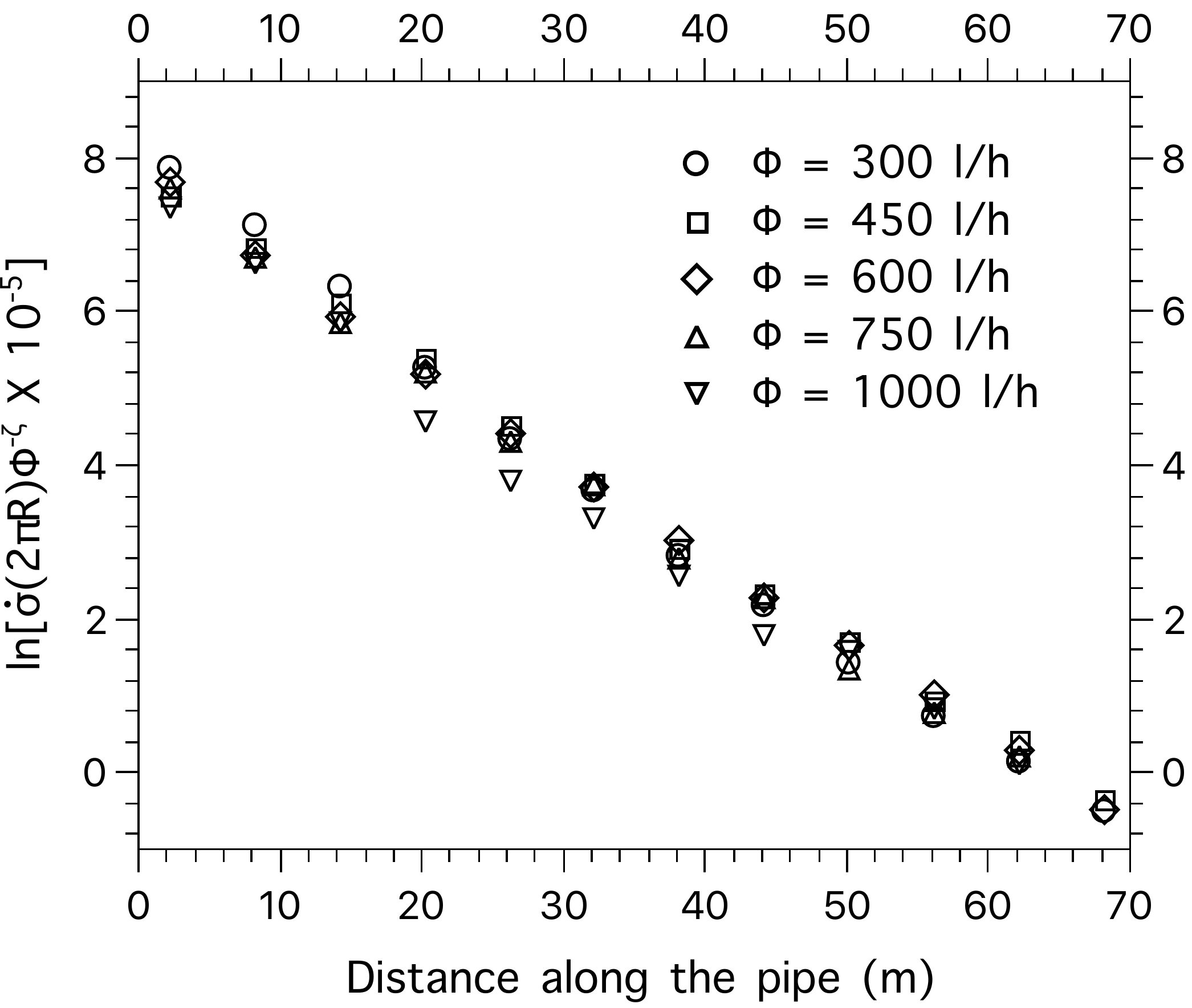}
\vspace{0.0cm}
\caption{A reasonable collapse of data for the mass deposition rates along all the pipe extension can be produced if they are normalized by a ``sliding" power of the flow rate, that is, $\Phi^\zeta$ with $\zeta = 1.0 + (i-1)/11$, where $i=1,2,...,12$ labels the sequence of measurement points.} 
\label{}
\end{figure}

We consider, initially, one of the flow cases where mass deposition rates were measured with the smallest relative error bars, say, the one with $\Phi = 300$ l/h and get $c$ straightforwardly from the mass deposition rate taken at the farthest measurement point downstream along the pipe. From the inspection of the several mass deposition rate curves, we also restrict to $x > 20$ m the range of positions where particulate deposition is supposed to dominate limescale formation. The pair of additional parameters, $a$ and $b$, are determined afterwards from optimizations based on the least square method, see Fig. 6. When dealing with other flow rate cases, $b$ and $c$ are not replaced by any other estimates. 
Optimal fittings are then established by adjustments of $a$, the only remaining free parameter in the application of {\hbox{Eq. (\ref{sigma_dot})}} to further experimental data. 

It turns out, from Eq. (\ref{sigma_infty}), that $\dot \sigma_\infty \Phi^{-2}$ is a flow rate independent constant $c$. In {\hbox{Fig. 7}}, the plots of $\dot \sigma(x) \Phi^{-2}$ for the several flow rate cases corroborates, in fact, the predicted collapse of data for larger values of the measurement position $x$. 

In Fig. 8, we show individual fittings of the measured mass deposition rates by the use of Eq. (\ref{sigma_dot}), obtained from optimal choices of $a$. The matching between analytical and experimental results is again very satisfactory.

To conclude the present analysis, we drive our attention to the regime of ion-by-ion limescale formation which is observed closer to the pipe's inlet. As stated in Eqs. (\ref{sigma_dot_ion}) and (\ref{sigma_dot}) (with $\alpha =0$),
the mass deposition rates are expected to crossover from a linear to a quadratic dependence on the flow rate, as measurement positions get farther away from the pipe's inlet. This piece of information is, actually, a useful clue to establish an extended collapse of data for the measured mass deposition rates. An educated guess is to introduce an interpolating scaling exponent $\zeta = \zeta(x)$, such that $\dot \sigma(x) \propto \Phi^{\zeta(x)}$, where $\zeta(x) = 1$ at the pipe's inlet and $\zeta(x) = 2$ at the pipe's outlet. Out of an infinity of possible interpolations, the simplest choice, the linear one, is found to yield a suggestive data collapse of the measured mass deposition rates for all of the investigated flow rate cases, as it can be clearly seen from Fig. 9. Results are reported there in monolog scale, since $\dot \sigma(x) \Phi^{- \zeta(x)}$ changes by a factor of $10^3$ as $x$ varies along the entire pipe's extension.

It is clear that the existence of a global data collapse like the one shown in Fig. 9 has great potential for practical applications. It yields a strong indication that limescale formation in industrial pipe flows could be mapped from the results of experimental tests performed at the scales of laboratory facilities.

\section{Conclusions and Outlook}

We have compiled, for the sake of model building, a number of phenomenological features of limescale formation in turbulent pipe flows. The microscopic characterization of the calcium carbonate scale, flow-induced pipe roughness, and particle size distributions were directly obtained from our experiments. Further important inputs came from current ideas on turbophoresis phenomenology. 

The top-down modeling strategy we have adopted places emphasis on general aspects of limescale formation and should eventually match results derived from the study of the detailed coupling between hydrodynamic phenomena and the chemical processes of particle nucleation and aggregation. It is reasonable to assume, from a heuristic point of view, that top-down and the more complex bottom-top approaches, like the one addressed in Kostoglou and Karabelas (1998), are absolutely fundamental to achieve progress in the understanding of limescale formation in the presence of turbulence flows.

The picture that emerges from our analyses is that both ion-by-ion and particulate deposition take place in large scale turbulent pipe flow experiments where the ionic solutions are mixed right at the pipe's inlet. We have been able to devise modeling expressions for the mass deposition rates related to those two distinct physical pictures, which are well supported by measurements.
Our main result, in short words, is the prediction that the mass deposition rate scales as \be
\dot \sigma \propto \frac{\Phi^\alpha}{R^\beta}  \ , \  \label{alpha_beta}
\ee
with $\alpha = 1$ and $\beta = 2$ for the case of ion-by-ion deposition, while $\alpha = 2$ and $\beta = 5$ are the scaling exponents predicted for the regime of particulate deposition.
Data collapse, as illustrated in Figs. 7 and 9, points out that a dynamic similarity theory of limescale formation is a promising perspective in the research horizon of particle-laden flows.

Much further work is in order. As an immediate task, it would be important to address the role of chemical concentrations and thermodynamic conditions in the behavior of the mechanical kinetic constants (\ref{par_a}), (\ref{par_b}), and (\ref{par_c}). Determinations of the scaling exponent $\beta$ described in (\ref{alpha_beta}) are also still completely open for experimental validation. 

Moving towards a deeper level of modeling, one would be interested to clarify the detailed mechanisms for the adhesion of the smaller calcium carbonate particles snapshotted in Figs. 2c and 2d, and the puzzlingly way how they are brought to the pipe's surface, as discussed in Sec. III. A plausible solution of this problem could be based on the (well reported) aggregation of metastable small vaterite globules, bonded into larger clusters 
(Rieger et al. 2007; Hu et al. 2012), which would then behave as inertial particles in the turbulent boundary layer.
\vspace{0,5cm}

\leftline{\it{Acknowledgments}}
\vspace{0.2cm}

The authors thank A. Barreto, D. Cruz,  E. Dittrich, D. Naiff, G. Pires, and F. Ramos for
enlightening discussions. This work has been partially supported by CNPq, CAPES and 
Petrobras (COPPETEC project number 20459).


\begin{appendix}

\section{Sizes of Surface-Adhering Particles}

Instead of digging into very specific details of the DLVO theory of microscopic particle-particle or 
particle-surface interactions (Israelachvili 1998), our aim, in this appendix, is just to get ``back of envelope" 
estimates for the sizes of calcium carbonate particles that can be deposited in wall-bounded turbulence. 
In contrast with electrochemical discussions which are carried out in equilibrium contexts, the following arguments 
take into account relevant dynamic properties of boundary layer flows. As a simplification, we focus our 
attention on calcium carbonate particles and surfaces which are electrically neutral.

To start, consider a spherical particle of radius $R$ which stands in the vicinity of a planar
material surface. Let $d$ be the distance between the material plane and the closest 
point on the particle's surface. The attractive van der Waals force between the particle and 
the surface can be written as (Israelachvili 1998)
\be
F_W(d) = \frac{A R }{6 d^2} \ , \
\ee
where $A$ is the so-called Hamaker constant,
\be
A = \pi^2 C n_1 n_2  \ , \  \label{hamaker}
\ee
and $n_1$ and $n_2$ represent the molecular number densities of the particle and the surface, while $C$ is a medium-dependent constant. For water, typically,
\be
C \simeq 1.4 \times 10^{-77} {\hbox{~J}} \cdot {\hbox{m}}^6 \ . \  \label{c-hamaker}
\ee
The whole point here is to understand whether a neutral spherical particle will be captured by the surface when transported by a turbulent flow. We expect to have particle adhesion if
\vspace{0.3cm}

(i) $F_W(d) > F_D  = 6 \pi \mu R v$, that is, if van der Waals forces overcome Stokes drag for the situation where $\mu$ is the dynamic viscosity of water
and $v$ is the fluid velocity in the lab's reference frame;
\vspace{0.2cm}

(ii) $\frac{1}{2} m v^2 < \int_d^\infty dx F_W(x) = AR/6 d$, that is, if a particle has enough kinetic energy to escape from the surface, once its departing (wall-normal) velocity at distance $d$ is $v$.
\vspace{0.3cm}

Conditions (i) and (ii) can be reshuffled, respectively, as 
\vspace{0.3cm}

(i') $f_1(d) \equiv A / (36 \pi d^2 \mu v) > 1$ and (ii') $f_2(d) \equiv A / (4 \rho_p R^2 d v^2) > 1$.
\vspace{0.3cm}

\vspace{-0.0cm}
\begin{figure}[h]
\begin{center}
\includegraphics[width=0.7\textwidth]{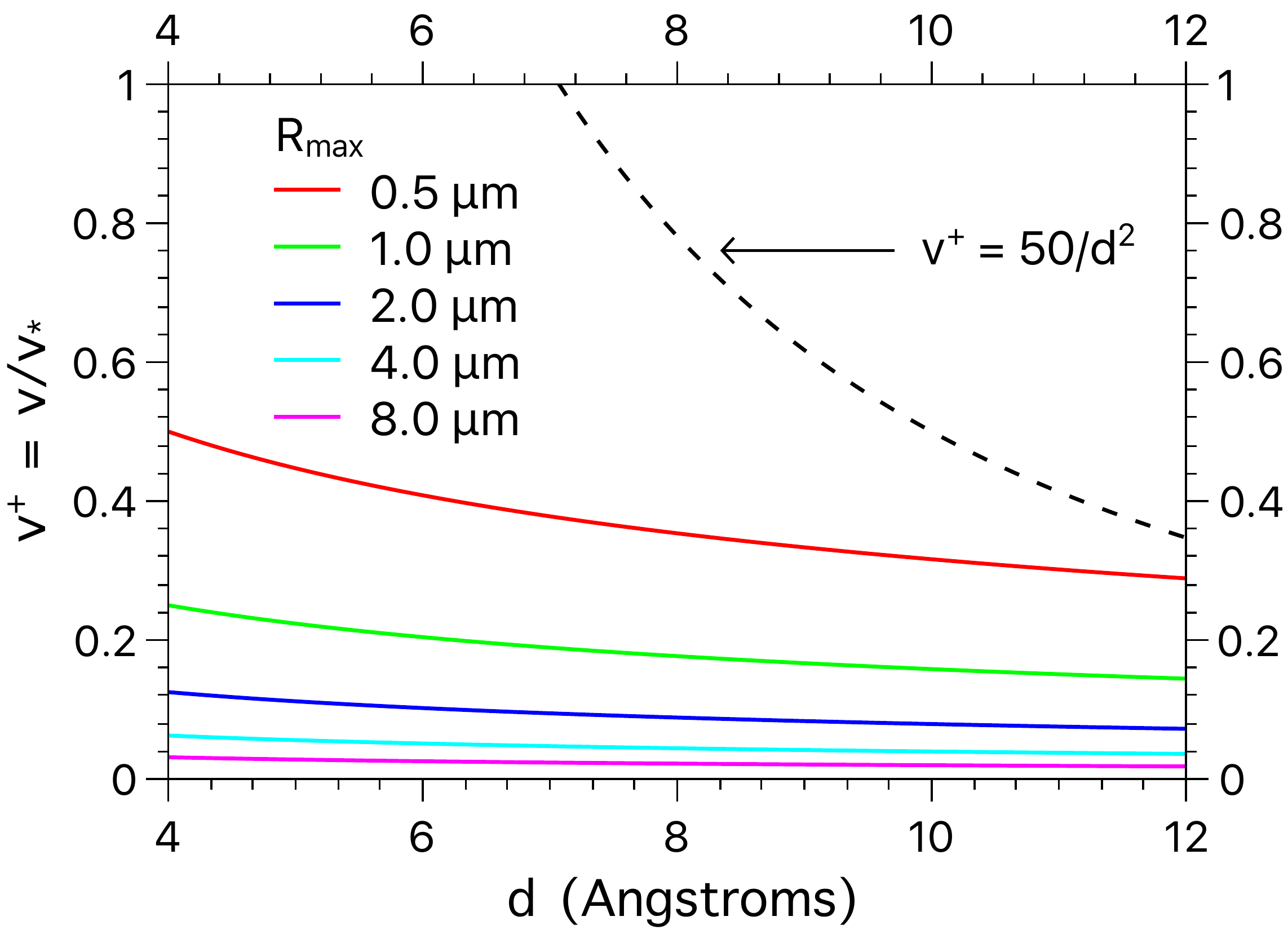}
\vspace{-0.0cm}
\caption{Values of $R_{max}$ for the flow rate case $\Phi = 600$ l/h with Re $\simeq 2 \times 10^4$. 
The colored curves indicate, for each pair of parameters, $(v/v_\ast,d)$, the values of maximum particle radius 
for electric capture (van der Waals) by the surface. Particle adhesion does not occur if $(v/v_\ast,d)$ 
lies above the dashed line.}
\label{vr}
\end{center}
\end{figure}

As a case study, we consider a pipe flow in its initial stages of limescale formation, such that surface roughness can be neglected, with Re = $2 \times 10^4$ (flow rate of $\simeq 600$ l/h in our experiments). The related friction Reynolds number is $R_\tau = 2400$ and the friction velocity is, accordingly, $v_\ast = \nu  R_\tau /D \simeq  0.2$ m/s. The calcium carbonate molecular number density and its mass density are, respectively, $n = 2.7 \times 10^{27}$/m$^3$ and $\rho_p = 2.7 \times 10^3$ Kg/m$^3$. For the interaction of calcium carbonate species, the Hamaker constant is evaluated, from (\ref{hamaker}) and (\ref{c-hamaker}), as $A \simeq 1.1 \times 10^{-20}$ J. 

For a particle with departing dimensionless velocity $v^+ \equiv v / v_\ast$ at a distance $d$, in Angstroms, from the attracting surface, condition (ii') leads
to the maximum particle radius for deposition,
\be
R_{max} \simeq \sqrt{\frac{1}{d (v^+)^2}} ~ \mu {\hbox{m}} \ . \
\ee
Condition (i'), on its turn, gives
\be
v^+ < \frac{50}{d^2}  \ . \ 
\ee
We get, from Fig. 10, graphic information on the values of $R_{max}$ associated to physically meaningful ranges of $d$ and $v^+$ in a turbulent boundary layer. This analysis suggests that transported neutral calcium carbonate particles can be deposited only if they have at most sizes of a few micrometers, in qualitative agreement with the SEM images reported in Figs. 2c and 2d.

\end{appendix}
\vspace{1.0cm}

\noindent {\bf{\large{References}}}
\vspace{1.0cm}


\hrule
\vspace{0.3cm}

\noindent Barabasi A-L, Stanley HE (1995) Fractal Concepts in Surface Growth. Cambridge University Press.
\vspace{0.3cm}

\hrule
\vspace{0.3cm}

\noindent Bello O (2017) Calcium Carbonate Scale Deposition Kinetics on Stainless Steel Surfaces. PhD Dissertation, The University of Leeds. https://etheses.whiterose.ac.uk/17647/
\vspace{0.3cm}

\hrule
\vspace{0.3cm}
\newpage
\hrule
\vspace{0.3cm}

\noindent Broby M, Neteland N, Ma X, Andreassen J-P, Seiersten M (2016) Scaling of Calcium Carbonate on Heated Surfaces - Crystallization or Particulate Fouling?. Proc. of the SPE International Oilfield Scale Conference and Exhibition, Aberdeen, Scotland, UK. \\
https://doi.org/10.2118/179901-MS.
\vspace{0.3cm}

\hrule
\vspace{0.3cm}

\noindent Burchette TP (2012) Carbonate rocks and petroleum reservoirs: A geological perspective from the industry.
Geol. Soc. Spec. Publ. 370(1): 17-37. \\
http://dx.doi.org/10.1144/SP370.14
\vspace{0.3cm}

\hrule
\vspace{0.3cm}

\noindent Cao Y, Xin M (2020) Numerical Simulation of Supercooled Large Droplet Icing Phenomenon: A Review.
Arch. Comput. Methods Eng. 27: 1231-1265. \\
https://doi.org/10.1007/s11831-019-09349-5
\vspace{0.3cm}

\hrule
\vspace{0.3cm}

\noindent Carino A, Testino A, Andalibi MR, Pilger F, Bowen P, Ludwig C (2017)
Thermodynamic-Kinetic Precipitation Modeling. A Case Study: The Amorphous Calcium Carbonate (ACC) Precipitation Pathway Unravelled. Cryst. Growth Des. 17(4): 2006-2015. \\ https://doi.org/10.1021/acs.cgd.7b00006
\vspace{0.3cm}

\hrule
\vspace{0.3cm}

\noindent Çengel YA, Cimbala (2006) Fluid Mechanics. McGraw-Hill.
\vspace{0.3cm}

\hrule
\vspace{0.3cm}

\noindent Cosmo RP, Pereira FAR, Ribeiro DC, Barros WQ, Martins AL (2019)
Estimating CO2 degassing effect on CaCO3 precipitation under oil well conditions.
J. Petrol. Sci. Eng. 181: 106207.
https://doi.org/10.1016/j.petrol.2019.106207
\vspace{0.3cm}

\hrule
\vspace{0.3cm}

\noindent de Souza AVA, Rosário F, Cajaiba J (2019)  
Evaluation of calcium carbonate inhibitors using sintered metal filter in a pressurized dynamic system. Materials 12(11): 1849 (2019). \\
https://doi.org/10.3390/ma12111849
\vspace{0.3cm}

\hrule
\vspace{0.3cm}

\noindent Gebauer D, V\"olkel A, C\"olfen H (2008) Stable Prenucleation Calcium Carbonate Clusters.
Science 322: 1819-1822. https://doi.org/10.1126/science.1164271
\vspace{0.3cm}

\hrule
\vspace{0.3cm}

\noindent Guha A (2008) Transport and Deposition of Particles in Turbulent and Laminar Flow.
Annu. Rev. Fluid Mech. 40: 311-341. https://doi.org/10.1146/annurev.fluid.40.111406.102220
\vspace{0.3cm}

\hrule
\vspace{0.3cm}
\newpage
\hrule
\vspace{0.3cm}

\noindent Hu Q, Zhang J, Teng H, Becker U (2012)
Growth process and crystallographic properties of ammonia-induced vaterite.
Am. Min. 97: 1437-1445. \\
https://doi.org/10.2138/am.2012.3983
\vspace{0.3cm}

\hrule
\vspace{0.3cm}

\noindent Israelachvili JN (1998) Intermolecular and Surface Forces. Academic Press Limited.
\vspace{0.3cm}

\hrule
\vspace{0.3cm}

\noindent Koishi A (2017) Carbonate mineral nucleation pathways.  PhD Dissertation, Université Grenoble Alpes. 
https://tel.archives-ouvertes.fr/tel-01701947/document
\vspace{0.3cm}

\hrule
\vspace{0.3cm}

\noindent Kostoglou M, Karabelas AJ (1998) Comprehensive Modeling of Precipitation and Fouling in Turbulent Pipe Flow.
Ind. Eng. Chem. Res. 37: 1536-1550.\\
https://doi.org/10.1021/ie970559g
\vspace{0.3cm}

\hrule
\vspace{0.3cm}

\noindent Koutsoukos PG, Kontoyannis CG (1984) Precipitation of calcium carbonate in aqueous solutions. J. Chem. Soc. Faraday Trans. 1 80: 1181-1192. \\ https://doi.org/10.1039/F19848001181 
\vspace{0.3cm}

\hrule
\vspace{0.3cm}

\noindent Kumar S, Naiya TK, Kumar T (2018) Developments in oilfield scale handling towards green technology-A review.
J. Petrol. Sci. Eng. 169: 428-444. \\ 
https://doi.org/10.1016/j.petrol.2018.05.068
\vspace{0.3cm}

\hrule
\vspace{0.3cm}

\noindent Lin L, Jiang W, Xu X, Xu P (2020) A critical review of the application of electromagnetic fields for scaling control in water systems: mechanisms, characterization, and operation. NPJ Clean Water 3: 25.
https://doi.org/10.1038/s41545-020-0071-9
\vspace{0.3cm}

\hrule
\vspace{0.3cm}

\noindent MacAdam J, Parsons SA (2004) Calcium carbonate scale formation and control. Rev Environ Sci Biotechnol 3: 159–169 (2004). https://doi.org/10.1007/s11157-004-3849-1
\vspace{0.3cm}

\hrule
\vspace{0.3cm}

\noindent Marchioli C, Soldati A (2002)  Mechanisms for particle transfer and segregation in a turbulent boundary layer. J. Fluid Mech. 468: 283-315. \\ https://doi.org/10.1017/S0022112002001738
\vspace{0.3cm}

\hrule
\vspace{0.3cm}

\noindent Mazumder MAJ (2020) A Review of Green Scale Inhibitors: Process, Types, Mechanism and Properties. 
Coatings 10(10): 928. https://doi.org/10.3390/coatings10100928
\vspace{0.3cm}

\hrule
\vspace{0.3cm}
\newpage
\hrule
\vspace{0.3cm}

\noindent Neville A (2012) Surface Scaling in the Oil and Gas Sector: Understanding the Process and Means of Management. Energ Fuel 26(7): 4158–4166. \\
https://doi.org/10.1021/ef300351w
\vspace{0.3cm}

\hrule
\vspace{0.3cm}

\noindent Nielsen MH, Aloni S, De Yoreo JJ (2014) In situ TEM imaging of CaCO3 nucleation 
reveals coexistence of direct and indirect pathways. Science 345: 1158-1162. \\
https://doi.org/10.1126/science.1254051
\vspace{0.3cm}

\hrule
\vspace{0.3cm}


\noindent Olajire AA (2015) A review of oilfield scale management technology for oil and gas production.
J.  Petrol. Sci. Eng. 135: 723-737. \\ https://doi.org/10.1016/j.petrol.2015.09.011
\vspace{0.3cm}

\hrule
\vspace{0.3cm}

\noindent Picano F, Sardina G, Casciola CM (2009)
Spatial development of particle-laden turbulent pipe flow.
Phys. Fluids 21: 093305.
https://doi.org/10.1063/1.3241992
\vspace{0.3cm}

\hrule
\vspace{0.3cm}

\noindent Pope SB (2000) Turbulent Flows. Cambridge University Press.
\vspace{0.3cm}

\hrule
\vspace{0.3cm}

\noindent Ramkrishna D (2000) Population Balances: Theory and Applications to Particulate Systems in Engineering.
Academic Press.
\vspace{0.3cm}

\hrule
\vspace{0.3cm}

\noindent Reeks MW (1983) The transport of discrete particles in inhomogeneous turbulence. \\
J. Aerosol Sci. 14: 729-739. 
https://doi.org/10.1016/0021-8502(83)90055-1
\vspace{0.3cm}

\hrule
\vspace{0.3cm}

\noindent Rieger J, Frechen T, Cox TG, Heckmann W, 
Schmidt C, Thieme J (2007)  
Precursor structures in the crystallization/precipitation processes of CaCO3 and control of particle formation by polyelectrolytes.
Faraday Discuss. 136: 265-277. \\
https://doi.org/10.1039/B701450C
\vspace{0.3cm}

\hrule
\vspace{0.3cm}

\noindent Sardina G, Schlatter P, Brandt L, Picano F, Casciola CM (2012) 
Wall accumulation and spatial localization in particle-laden wall flows.
J. Fluid Mech. 699: 50-78. \\
https://doi.org/10.1017/jfm.2012.65
\vspace{0.3cm}

\hrule
\vspace{0.3cm}
\noindent Vetter OJ, Farone WA (1987) Calcium Carbonate Scale in Oilfield Operations. Paper presented at the SPE Annual Technical Conference and Exhibition, Dallas, Texas, September. doi: https://doi.org/10.2118/16908-MS
\vspace{0.3cm}

\hrule
\vspace{0.3cm}
\newpage

\hrule
\vspace{0.3cm}

\noindent Wallace AF, Hedges LO, Fernandez-Martinez A, Raiteri P, 
Gale JD, Waychunas GA, Whitelam S, Banfield JF, De Yoreo  JJ (2013) 
Microscopic Evidence for Liquid-Liquid Separation in Supersaturated CaCO$_3$ Solutions. Science 341: 885-889. \\
https://doi.org/10.1126/science.1230915
\vspace{0.3cm}

\hrule
\vspace{0.3cm}

\noindent Witten TA, Sander LM (1983)  Diffusion-limited aggregation. 
Phys. Rev. B 27: 5686-5697.
https://doi.org/10.1103/PhysRevB.27.5686
\vspace{0.3cm}

\hrule
\vspace{0.3cm}

\noindent Young J, Leeming A (1997) A theory of particle deposition in turbulent pipe flow.  
J. Fluid Mech. 340: 129-159. https://doi.org/10.1017/S0022112097005284
\vspace{0.3cm}

\hrule
\vspace{0.3cm}


\end{document}